# Analysis of the Driving Factors of Implementing Green Supply Chain Management in SME's in the City of Semarang

[1]Nanang Adie Setyawan, [2]Hadiahti Utami, [3]Bayu Setyo Nugroho,
[4]Mellasanti Ayuwardi, [5]Suharmanto

[12345] Departement of Business Administration, Semarang State Polytechnic, Indonesia.

*Abstract:* This study set out to determine what motivated SMEs in Semarang City to undertake green supply chain management during the COVID-19 and New Normal pandemics. The purposive sampling approach was used as the sampling methodology in this investigation. There are 100 respondents in the research samples. The AMOS 24.0 program's structural equation modelling (SEM) is used in this research method. According to the study's findings, the Strategic Orientation variable significantly and favourably affects the Green Supply Chain Management variable expected to have a value of 0.945, and the Government Regulation variable has a positive and strong influence on the variable Green Supply Chain Management with an estimated value of 0.070, the Green Supply Chain Management variable with an estimated value of has a positive and significant impact on the environmental performance variable. 0.504, the Strategic Orientation variable with an estimated value of has a positive and significant impact on the environmental performance variable. 0.442, The Environmental Performance variable is directly impacted positively and significantly by the Government Regulation variable, with an estimated value of 0.041. This significant positive influence is because SMEs in Semarang City have government regulations, along with government support for facilities regarding efforts to implement the concept of environmental concern, causing high environmental performance caused by the optimal implementation of Green supply chain management is built on a collaboration between the government and the supply chain's participants.

## I. INTRODUCTION

The development of the business industry is currently increasingly stringent due to globalization and the rapid development of information technology. Competition between business actors is becoming increasingly stringent, business actors are expected to be able to increase their competitive advantage, including by increasing the integration and exchange of information between organizations and effective business processes along the Supply Chain Management is used in the supply chain. [1]. A technique for controlling the movement of goods, information, and money from upstream to downstream involves a number of partners, including suppliers, factories, distribution networks, and logistical services. [2].

Business actors must evaluate and employ green supply chain management due to rising demands and challenges in the economic and environmental rivalry. This aims to boost the financial rewards for business actors while also maintaining environmental sustainability. [3]. Environmental issues have also become one of the conversations of business actors, this is caused by increasing public awareness regarding environmentally friendly products. With public awareness encouraging business actors to implement environmentally friendly concepts in their businesses, therefore the current concept of environmentally friendly or Green Supply Chain Management cannot be underestimated by business actors [4].

Growing industrialisation and globalisation increase business prospects while also putting additional stress on the environment. All phases of a product's life cycle, including resource extraction, manufacturing, reuse, recycling, and disposal, have an impact on the environment. [5]. Practices of green supply chain management, such as green manufacturing, green distribution/marketing, and reverse logistics, refer to the engagement of [6].

From the extraction of raw materials to product design, manufacturing procedures, transportation of the finished product to consumers, and end-of-life management, supply chain management incorporates environmental considerations. [7]. As a result, GSCM has become a successful strategy for lowering the risk of environmental harm and the environmental load associated with manufacturing and disposing products while also increasing profitability and competitive advantage. [4].

The increase in business activity in addition to increasing profits for companies and the country's economy also has an impact on several losses, such as increased waste, pollution, and other pollution to the environment. So the environmental priority issues for the area of Central Java Province are determined as follows:

1. Decrease in Water Quality
2. Garbage and Hazardous Waste Management
3. Climate change
4. Land Use Change and Land Degradation
5. Coastal Zone Management

One of the business sectors that needs handling regarding environmental impacts is the traditional market because the market is a point where SMEs' businesses gather, so it has a high contribution to environmental pollution which causes low environmental performance [8]. This research wants to test and explore more about the environmental impacts contributed by SMEs in the city of Semarang. Previous years of research and service on Fashion SMEs and ornamental plant SMEs in cities/districts. Semarang that sharing knowledge, can contribute to improving employee and organizational performance which will certainly have an impact on increasing productivity and sales or organizational profits [9]

Low environmental performance requires the application of a broader concept of sustainability, while the concept of sustainability itself leads to the expansion of green policies and standards that cover the entire supply chain. The application of GSCM in the business sector requires driver factors or pressure factors both from internal and external parties of the company because the pressure factor will trigger good and sustainable implementation, research conducted by resulted in findings that the most effective pressure from implementing GSCM was strategic orientation and government regulation [1].

The research was conducted in the MSE`S sector in the city of Semarang, on the other hand, there is a problem where there are several piles of garbage around MSE`S in the Semarang area so it would be better if MSE`S implemented the GSCM system to reduce MSE`S wasteWhen examining the experiences of industrialised nations on the function and contribution of small and medium-sized firms (SMEs) in the nation's economic growth, developing countries are now beginning to shift their approach. The Semarang city government is one of the regions in Central Java province that pays attention to the problem of SMEs in its area. Therefore, it would be better if SMEs in Semarang pay attention to the GSCM sector to increase the effectiveness of production efficiency and to preserve the surrounding natural environment [5].

Based on the elaboration of the background of the previous problem, the problem of this research is " How to Influence the Driving Factors for the Implementation of Green SupplyChain Management for SMEs in Semarang City "

## II. MATERIALS AND METHODS

The planned research time requirement from the initial stage to the end is 6 (six) months. Place or location of the research survey in Semarang City, where the research survey area consists of 16 sub-districts. where primary data were used in this investigation. The responses (perceptions) of respondents regarding strategic orientation, governmental regulation, green supply chain management, and environmental performance constitute the study's main source of data. Data was collected using a survey method in the form of a questionnaire. To get interval data, the statements in this questionnaire are made using a scale of 1–5 Semantic Differential, and they are assigned a score or value.

SME businesses in Semarang make up the study's sample and population. Purposive sampling was the method of sampling that was employed in this investigation. The requirements for the sample are first, SMEs in Semarang City both offline and online; Second, respondents can be found and are willing to fill out a questionnaire. The number of samples in this study refers to where the minimum sample is 100 for the SEM analysis tool [10]. Model and hypothesis testing are carried out by Regression Equation Test which is divided into Confirmatory factor analysis, Regression Weight in Structural Equation Modeling (SEM), which is used to assess the strength of the relationship between variables, and Structural Equation Modeling (SEM), which employs the AMOS 24.0 computerised programme in this case to confirm the most dominating factors in a set of variables. [11].

## III. RESULTS AND DISCUSSION

*A. Characteristics Respondents*

Table 1 shows the characteristics of the respondents as subjects in the study.

**Table 1: Characteristics of Respondents**

| Criteria | Characteristics | Percentage |
|---|---|---|
| Gender | Woman | 70 % |
|  | Man | 30 % |
| Education | Elementary School | 3 % |
|  | Junior High School | 2 % |
|  | Senior High School | 51 % |
|  | Diploma | 14 % |
|  | Bachelor | 29 % |
|  | Master | 1 % |
| SME business age | < 3 years | 54 % |
|  | 3-5 years | 16 % |
|  | 6-8 years | 7 % |
|  | > 8 years | 23 % |
| Business Category | Culinary/Restaurant/Cafe | 56 % |
|  | Agribusiness | 1 % |
|  | Creative Products | 5 % |
|  | Trade Services | 22 % |
|  | Industry/Manufacturing | 3 % |
|  | Other | 13 % |

*Source: Primary data processed, 2022.*

*B. Data Normality Evaluation*

The findings of the normalcy test indicated that the data were normally distributed univariately and multivariate with no univariate value exceeding the critical limit (cr) of a variable of ±2.58 and in a multivariate presentation of -0.606. Processed data can be said to be normal if it has a critical value (cr) which is at ± 2.58 and the results of the univariate and multivariate data normality tests show that the value is still in the vulnerable value of ±2.58 [12].

*C. Univariate & Multivariate Outlier Evaluation*

Mahalanobis Distance is used to measure the presence or absence of outlier data (damaging data), namely by looking at the observation scores that are very different from the centroid scores for 100 cases. Based on the Mahalanobis distance, the minimum distance listed for Mahalanobis is 9.752 and the maximum distance is 32,498. The outlier data is perceived from the Mahalanobis value which exceeds the chi-square value. In this study, the chi-square of 19 degrees of freedom (number of variable indicators) at a significance level of 0.01, namely 32,852, indicates that there are no outliers [12].

*D. Multicollinearity Evaluation*

Multicollinearity symptoms can be seen through matrix sample correlations, if the resulting value of each indicator is smaller than (<) 0.90 then it can be stated that there are no multicollinearity symptoms. In this study, the results of data processing showed that there were no symptoms of multicollinearity in the matrix sample correlations of 19 indicators spread across the six variables tested [12].

*E. Measurement Model Test*

In this study, it can be seen that the chi-square value ($X^2$) and the degree of freedom (df) value. As may be observed from the writing model test results, that chi-square ($X^2$) has a value of 158.919 and the degree of freedom (df) has a value of 109.

**Table 2: Evaluation Results *Cut Value* Criteria**

| *Goodness-of-fit* index | *Cut of Value* | Analysis Results | Model Evaluation |
|---|---|---|---|
| Chi-Square | ( Small ) ≤ 160. 750 | 158,919 | Well |
| probability | ≥ 0.05 | 0.061 | Well |
| GFI | ≥ 0.90 | 0.870 | Marginal |

| TLI | ≥ 0.90 | 0.969 | Well |
|---|---|---|---|
| CFI | ≥ 0.90 | 0.981 | Well |
| DF /CMIN | ≤ 2.00 | 1,458 | Well |
| RMSEA | ≤ 0.08 | 0.068 | Well |

*Source: Primary data processed, 2022.*

The results of the model test are shown in Figure 1 and highlight the AMOS 24 program's goodness of fit standards. It shows that the structural equation modeling analysis in this study is acceptable according to the fit model with a Chi-square value = 158, 919, Probability = 0.061 DF/CMIN = 1.458, GFI = 0.870, CFI = 0.981, TLI = 0.981 and RSMEA = 0.068. It is clear from this fit model that the model satisfies the goodness of fit requirements. Therefore the structural equation model in this study is suitable and feasible to use so that it can be interpreted for further discussion [12].

Below is a picture of the results of the analysis in this study which includes the following variables: Strategic orientation variable with five (4) indicators, Government Regulation with (2) two indicators, Green Supply Chain Management with (6) six indicators, Environmental Performance with (7) seven indicators, shown in Figure 1 as follows:

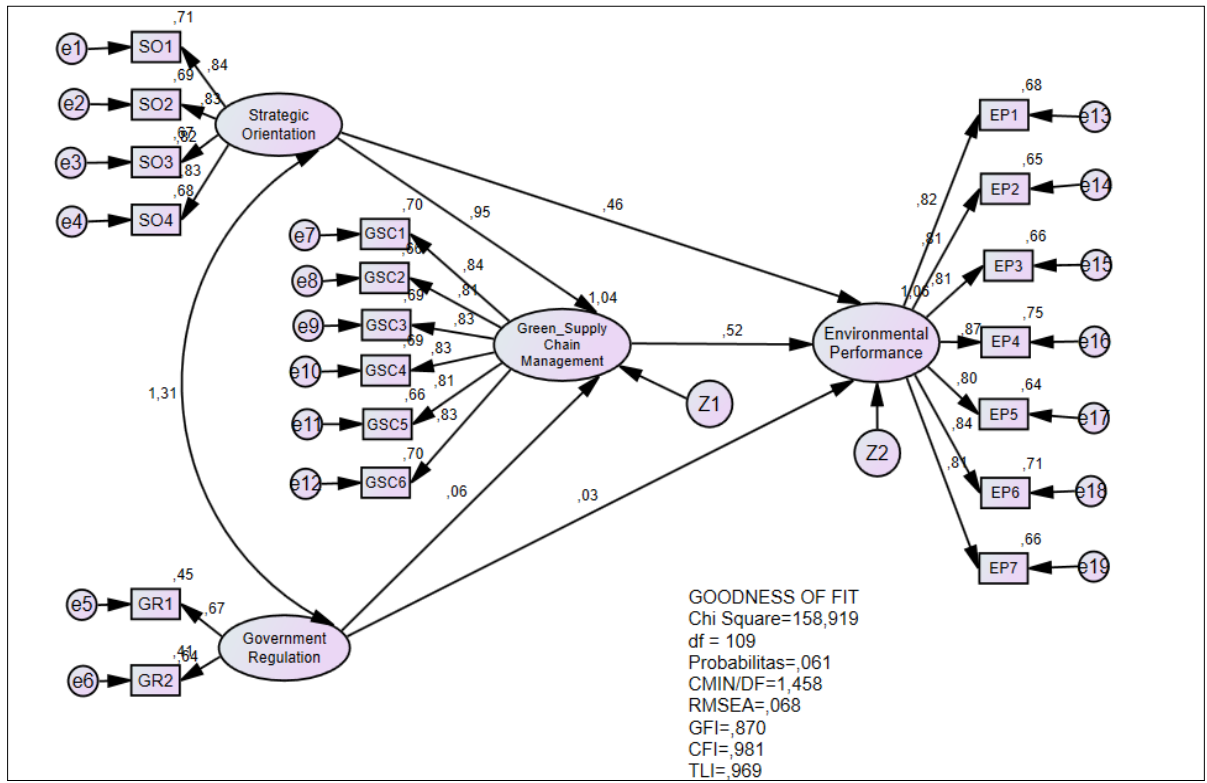

*Source: Primary data processed, 2022.*
**Figure 1. Results of analysis of the research structural model**

### F. Test hypothesis
In the testing phase of the hypothesis of a significant causal relationship, the value of the critical ratio (cr) has a critical T value of ≥ 1.966. In facilitating decision-making, the authors can see from the probability figure (P) where (P) ≤ 0.05. If the P value ≤ 0.05 then H $_{0\ is}$ accepted, and if vice versa if the P value ≥ 0.05 then H $_{0\ is}$ rejected, in the SEM package with the A mos 24 application the results of the hypothesis test can be seen through the output regression weights. (Gh ozali, 2017). Table 3 displays the findings of the hypothesis test.

**Table 3: Hypothesis Test Results (Regression Weights)**

| Endogenous Variables | | Exogenous Variables | Estimates | SE | CR | P |
|---|---|---|---|---|---|---|
| Green Supply Chain Management | <--- | Strategic Orientation | .945 | .084 | 11.202 | *** |
| Green Supply Chain Management | <--- | Government Regulations | .070 | .031 | 2,254 | .024 |
| Environmental Performance | <--- | Green Supply Chain Management | .504 | .188 | 2,685 | .007 |
| Environmental Performance | <--- | Strategic Orientation | .442 | .181 | 2,443 | .015 |
| Environmental Performance | <--- | Government_Regulation | .041 | .013 | 3.101 | .002 |
| GSC2 | <--- | Green Supply Chain_Management | .856 | .096 | 8,874 | *** |
| EP1 | <--- | Environmental Performance | 1,000 | | | |
| EP2 | <--- | Environmental Performance | .947 | .113 | 8,400 | *** |
| EP3 | <--- | Environmental Performance | .960 | .093 | 10,320 | *** |
| GSC1 | <--- | Green Supply Chain_Management | 1,000 | | | |
| GSC3 | <--- | Green Supply Chain Management | .934 | .088 | 10,660 | *** |
| GR1 | <--- | Government Regulations | 1,000 | | | |
| GSC4 | <--- | Green Supply Chain_Management | 1.016 | .095 | 10,681 | *** |
| GSC5 | <--- | Green Supply Chain_Management | .919 | .089 | 10.273 | *** |
| GSC6 | <--- | Green_Supply_Chain_Management | 1,047 | .098 | 10,730 | *** |
| EP4 | <--- | Environmental_Performance | 1,140 | .100 | 11,369 | *** |
| EP5 | <--- | Environmental_Performance | .883 | .088 | 10,086 | *** |
| EP6 | <--- | Environmental_Performance | 1,087 | .086 | 12,704 | *** |
| EP7 | <--- | Environmental_Performance | .897 | .087 | 10,278 | *** |
| GR2 | <--- | Government_Regulation | .882 | .113 | 7,805 | *** |
| SO4 | <--- | Strategic_Orientation | .955 | .088 | 10,900 | *** |
| SO3 | <--- | Strategic_Orientation | .884 | .084 | 10,551 | *** |
| SO2 | <--- | Strategic_Orientation | .978 | .109 | 9.005 | *** |
| SO1 | <--- | Strategic_Orientation | 1,000 | | | |

*Source: Primary data processed, 2022.*

The output results on Regression Weights present that each indicator or manifest variable that reflects the latent variable has a critical ratio (CR) value greater ( > ) than 1.96, the same as the t value in regression ( > ) 1.96 and P (Probability of significance) with P < 0.05, it can be concluded that the six hypotheses are accepted.

The elaboration of the output results on Regression Weights is as follows. The Strategic Orientation variable, with an estimated value of 0.945, significantly and positively influences the Green Supply Chain Management variable, the Government Regulation variable has a large and favourable impact on the Green Supply Chain Management variable with an estimated value of 0.070, the Green Supply Chain Management variable has a strong and positive impact influence on the Environmental Performance variable with an estimated value of 0.504, With an estimated value of 0.442, the Strategic Orientation variable has a substantial positive relationship with the Environmental Performance variable, the Government Regulation variable has a direct positive and significant influence on the Environmental Performance variable with an estimated value of 0.041. In this case, it can be explained as follows.

The Green Supply Chain Management strategy orientation variable and the correlation coefficient have a substantial positive association of 0.945 or 94%, which means that when the strategy orientation variable is increased by one time, the Green Supply Chain Management variable will increase also by 94% , Government Regulation Variable or Government Regulations and Green Supply Chain Management Have a Significantly Positive Relationship with a correlation coefficient of 0.07 or 7%, which means that when the government regulation variable is increased by one time, the Green Supply Chain variable Management will also increase by 7%, Green Supply Chain Management factors and environmental performance, or firm environmental performance, have a considerable positive association. with a correlation coefficient of 0.504 or 50.4%, which means that when the Green Supply Chain Management variable is increased by one time, the environmental performance variable will also increase of 50.4% , Strategic orientation characteristics and environmental performance as measured by Green Supply Chain Management are significantly positively correlated with a correlation coefficient of 0.442 or 44.2% where performance in terms of the environment is affected by the creation of a strategic commitment to green supply chain management. The SMEs in Semarang City are responsible for this large beneficial impact.

Their strategic orientation, willingness, and understanding of environmental concerns have led to high environmental performance as a result of the ongoing adoption of green supply chain management. optimal. Government regulation variables and environmental performance as measured by green supply chain management are significantly positively correlated with a correlation coefficient of 0.041 or 4.1%, where in this study Green Supply Chain Management acts as a perfect mediating variable. the findings of a significant positive relationship are shown in research that examines the direct effect between government regulation and environmental performance The government's regulations and support for facilities in Semarang City's efforts to implement the idea of environmental awareness have had a significant positive impact. High environmental performance is the result of the implementation of an ideal green supply chain management strategy, which is carried out based on the collaboration of supply chain participants and the government itself.

## IV. CONCLUSION

The factors of strategic orientation, governmental regulation, green supply chain management, and environmental performance are all included in this study. The results of testing the model using the AMOS 24 program show that the structural equation modeling analysis in this study is acceptable according to the fit model with a Chi-square value = 158, 919, Probability = 0.061 DF /CMIN = 1.458, GFI = 0.870, CFI = 0.981, TLI = 0.981 and RSMEA = 0.068. It is clear from this fit model that the model satisfies the goodness of fit requirements.

According to the results of testing the 5 hypotheses proposed in this study, it was concluded that the 5 hypotheses were accepted, namely the Strategic Orientation variable has a favourable and considerable impact on the variables. Manage a green supply chain with an estimated value of 0.945, variable Government Regulations has a favourable and considerable impact on the variables. Manage a green supply chain with an estimated value of 0.070, Variables are positively and significantly influenced by green supply chain management. Performance in the Environment 0.504, Variables are positively and significantly influenced by green supply chain management. Performance in the Environment with an estimated value of 0.442, variable Government regulation directly affects variables in a positive and meaningful way Environmental Performance with an estimated value of 0.041.

Future research needs to consider the object of research on other SMEs in various cities in Indonesia. In addition, it is necessary to collaborate on other variables related to the characteristics of the Indonesian people/population, including Spiritual Intelligence . In addition, it can be developed to influence the Business Performance of Small and Medium Enterprises. Supply chain management can also deepen its influence on business conditions in various SMEs in the city of Semarang, especially for manufactured products that involve processes from upstream to downstream. Such studies are needed to cross-validate research findings.